\documentclass[sigconf]{acmart}
\usepackage{graphicx} 
\usepackage{listings}
\usepackage{arydshln}
\usepackage{subcaption}
\usepackage{hyperref}
\usepackage{wrapfig}
\title{Towards Fuzzing Zero-Knowledge Proof Circuits \\
(Short Paper)}
\author{Stefanos Chaliasos}
\affiliation{%
  \institution{Imperial College London \& ZKSecurity}
  \country{United Kingdom}
}

\author{Imam Al-Fath}
\affiliation{%
  \institution{ZKSecurity}
  \country{USA}
}

\author{Alastair F. Donaldson}
\affiliation{%
  \institution{Imperial College London}
  \country{United Kingdom}
}

\newcommand{\empirical}[1]{#1}
\newcommand{\point}[1]{\par\smallskip\noindent\textbf{#1.}}

\copyrightyear{2025}  \acmYear{2025}  \setcopyright{cc}\setcctype{by}\acmConference[ISSTA Companion '25]{34th ACM SIGSOFT International Symposium on Software Testing and Analysis}{June      25--28, 2025}{Trondheim, Norway}\acmBooktitle{34th ACM SIGSOFT International Symposium on Software Testing and Analysis (ISSTA Companion '25), June      25--28, 2025, Trondheim, Norway}\acmDOI{10.1145/3713081.3731718}\acmISBN{979-8-4007-1474-0/2025/06}

\ccsdesc[500]{Security and privacy~Software security engineering}
\ccsdesc[500]{Security and privacy~Formal methods and theory of security}

\begin{document}

\begin{abstract}
Zero-knowledge proofs (ZKPs) have evolved from a theoretical cryptographic concept into a powerful tool for implementing privacy-preserving and verifiable applications without requiring trust assumptions.
Despite significant progress in the field, implementing and using ZKPs via \emph{ZKP circuits} remains challenging, leading to numerous bugs that affect ZKP circuits in practice,
and \emph{fuzzing} remains largely unexplored as a method to detect bugs in ZKP circuits. 
We discuss the unique challenges of applying fuzzing to ZKP circuits, examine the oracle problem and its potential solutions, and propose techniques for input generation and test harness construction. We demonstrate that fuzzing can be effective in this domain by implementing a fuzzer for \texttt{zk-regex}, a cornerstone library in modern ZKP applications. In our case study, we discovered \empirical{\textit{$13$}} new bugs that have been confirmed by the developers.
\end{abstract}

\keywords{Fuzzing, Zero-Knowledge Proofs}

\maketitle

\section{Introduction}
Zero-knowledge proofs (ZKPs) are cryptographic protocols that enable a prover to convince a verifier that a given statement is true without revealing any additional information~\cite{goldwasser1989knowledge}. They are widely used in applications ranging from privacy-preserving transactions and confidential voting to secure identity verification and blockchain scalability~\cite{ernstberger2024you}. To implement a ZKP, the statement must be transformed into an arithmetic circuit or constraint system. During the past decade, various domain-specific languages (DSLs) have been developed to assist developers in writing ZKP circuits~\cite{sheybani2025zero}.

Despite their promise, ZKPs are challenging to implement correctly. Errors in circuit design and implementation can compromise the security of a ZKP~\cite{chaliasos2024sok}. Such bugs may lead to issues with completeness (failure to produce proofs for valid statements), correctness (proofs that do not accurately reflect the intended computation), or soundness (dishonest provers generating invalid proofs that are nevertheless accepted by the verifier), thereby undermining system reliability. To address these challenges, existing solutions have primarily focused on formal verification~\cite{pailoor2023automated,coglio2023formal,miguel2024scalable} and static analysis~\cite{soureshjani2023automated,wen2023practical} to check the correctness of ZKP circuits. While these methods have contributed significantly to identifying vulnerabilities, formal verification tools often struggle with scalability when applied to complex circuits, and static analysis focuses on specific patterns while facing challenges in minimizing false positives~\cite{zkbugs}.

Given these limitations, fuzzing emerges as a promising complementary approach for testing ZKP circuits. 
Yet, applying fuzzing to ZKP circuits introduces unique challenges, such as designing ZKP-specific test oracles and generating appropriate witnesses, necessitating further research.

\point{Contributions} In this paper, we make the following contributions:
\begin{itemize}
    \item We present a comprehensive overview of the key challenges involved in fuzzing ZKP circuits.
    \item We sketch \emph{initial solutions}, to be further explored by future work, to tackle the test oracle problem and the complexities of input (witness) generation for fuzzing ZKP circuits.
    \item We identify and discuss several open problems and future research directions in the domain of fuzzing ZKP circuits, highlighting opportunities for further enhancement.
    \item We demonstrate the practical impact of fuzzing ZKP circuits by implementing a simple yet powerful fuzzer for a state-of-the-art ZKP library, which uncovered \empirical{$13$} new bugs.
\end{itemize}

\section{Background}
\label{sec:background}

\begin{figure*}[t]
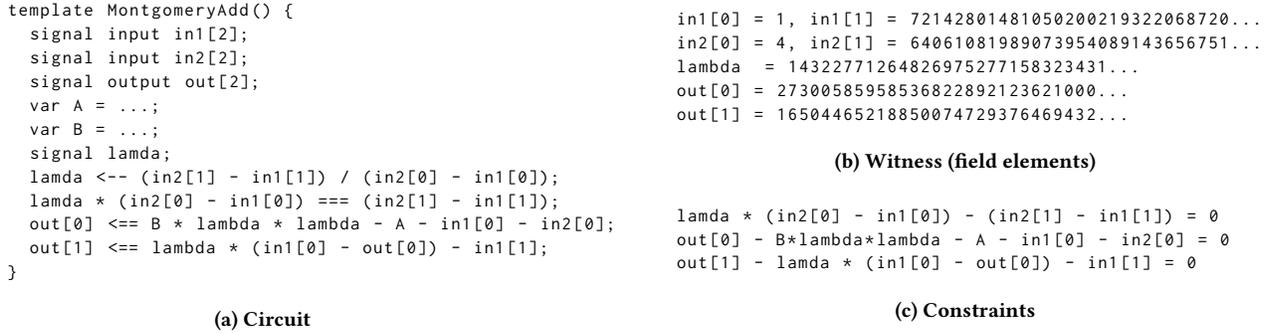

  \begin{subfigure}[c]{.45\linewidth}
\centering
    \begin{lstlisting}[basicstyle=\footnotesize\ttfamily]
    template MontgomeryAdd() {
      signal input in1[2];
      signal input in2[2];
      signal output out[2];
      var A = ...;
      var B = ...;
      signal lamda;
      lamda <-- (in2[1] - in1[1]) / (in2[0] - in1[0]);
      lamda * (in2[0] - in1[0]) === (in2[1] - in1[1]);
      out[0] <== B * lambda * lambda - A - in1[0] - in2[0];
      out[1] <== lambda * (in1[0] - out[0]) - in1[1];
    }
    \end{lstlisting}
    \caption{Circuit}
    \label{fig:montgomery-circuit}
  \end{subfigure}\hfill
  \begin{tabular}[c]{@{}c@{}}
    \begin{subfigure}[c]{.50\linewidth}
    \centering
    \begin{lstlisting}[basicstyle=\footnotesize\ttfamily]
    in1[0] = 1, in1[1] = 72142801481050200219322068720...
    in2[0] = 4, in2[1] = 64061081989073954089143656751...
    lambda  = 14322771264826975277158323431...
    out[0] = 27300585958536822892123621000...
    out[1] = 16504465218850074729376469432...
    \end{lstlisting}
    \caption{Witness (field elements)}
    \label{fig:montgomery-witness}
    \end{subfigure}\\
    \noalign{\bigskip}%
    \begin{subfigure}[c]{.50\linewidth}
    \centering
    \begin{lstlisting}[basicstyle=\footnotesize\ttfamily]
    lamda * (in2[0] - in1[0]) - (in2[1] - in1[1]) = 0
    out[0] - B*lambda*lambda - A - in1[0] - in2[0] = 0
    out[1] - lamda * (in1[0] - out[0]) - in1[1] = 0
    \end{lstlisting}
    \caption{Constraints}
    \label{fig:montgomery-constraints}
    \end{subfigure}
  \end{tabular}
  \caption
    {MontgomeryAdd circuit, with a witness sketch and the constraints generated from the circuit}\label{fig:montgomeryoverall}
\end{figure*}
\begin{figure*}[t]
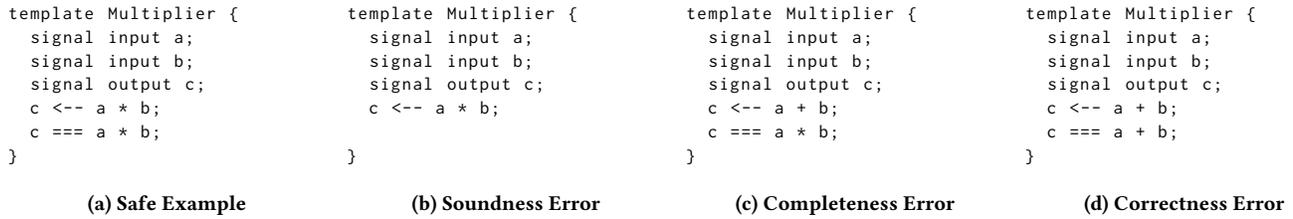

\centering
\begin{subfigure}[b]{0.24\textwidth}
\centering
\begin{lstlisting}[basicstyle=\footnotesize\ttfamily]
template Multiplier {
  signal input a;
  signal input b;
  signal output c;
  c <-- a * b;
  c === a * b;
}
\end{lstlisting}
\caption{Safe Example}
\label{fig:multiplierSafe}
\end{subfigure}
\hfill
\begin{subfigure}[b]{0.24\textwidth}
\centering
\begin{lstlisting}[basicstyle=\footnotesize\ttfamily]
template Multiplier {
  signal input a;
  signal input b;
  signal output c;
  c <-- a * b;

}
\end{lstlisting}
\caption{Soundness Error}
\label{fig:multiplierSoundness}
\end{subfigure}
\hfill
\begin{subfigure}[b]{0.24\textwidth}
\centering
\begin{lstlisting}[basicstyle=\footnotesize\ttfamily]
template Multiplier {
  signal input a;
  signal input b;
  signal output c;
  c <-- a + b;
  c === a * b;
}
\end{lstlisting}
\caption{Completeness Error}
\label{fig:multiplierCompleteness}
\end{subfigure}
\hfill
\begin{subfigure}[b]{0.24\textwidth}
\centering
\begin{lstlisting}[basicstyle=\footnotesize\ttfamily]
template Multiplier {
  signal input a;
  signal input b;
  signal output c;
  c <-- a + b;
  c === a + b;
}
\end{lstlisting}
\caption{Correctness Error}
\label{fig:multiplierCorrectness}
\end{subfigure}
\caption{Four Circom multiplier examples, illustrating a correct circuit and various kinds of error
}
\label{fig:fourCircomExamples}
\end{figure*}

A zero-knowledge proof (ZKP) is a cryptographic protocol by which a prover convinces a verifier that a given statement is true without revealing anything else beyond the truth of the statement~\cite{goldwasser1989knowledge}. A valid ZKP must satisfy three properties: (i) \textit{completeness}, if the statement is true and both parties are honest, the verifier will accept the proof; (ii) \textit{soundness}, if the statement is false, no cheating prover can convince the verifier to accept the proof; and (iii) \textit{zero-knowledge (ZK)}, the verifier learns nothing beyond the fact that the statement is true. In practice, many modern ZKPs are non-interactive and succinct. For example, zkSNARKs (zero-knowledge succinct non-interactive arguments of knowledge)~\cite{gennaro2013quadratic} allow very short proofs and fast verification, making them practical for applications like blockchain and privacy-preserving protocols~\cite{ernstberger2024you}. These protocols require the statement to be expressed as an NP-hard problem (usually as an arithmetic circuit or constraint system)~\cite{ernstberger2023zk}. ZKPs are widely used for \emph{private computations} (e.g., shielding transaction details in Zcash~\cite{sasson2014zerocash}) and for \emph{verifiable computations} (e.g., achieving blockchain scalability with validity rollups~\cite{chaliasos2024analyzing}). Note that in verifiable computation, the \textit{ZK property is not required}. Instead, the key advantage is that a verifier can accept a succinct proof of a potentially large or expensive computation using significantly fewer resources than re-running the entire computation. For an in-depth introduction to ZKPs, we refer the reader to~\citet{thaler2022proofs}.

\textbf{ZKP Circuits.}
To obtain a ZKP of a statement, a developer must represent that statement in the form of a circuit. In particular, a \emph{ZKP circuit} is an arithmetic representation of the computation, modeled as a collection of polynomial equations (constraints) over a finite field~\cite{gennaro2013quadratic}. These constraints define precisely which values the circuit will accept.

A circuit $C$, defines a relation $R(I, W)$, where $I$ is a public input provided by the verifier, and $W$ is a witness computed by the prover that satisfies $R$.
For instance, given a function, $f$ over public input $x$, consider the problem of proving that  $y=f(x)$ where $y$ is claimed to be an output computed by $f$.
The constraints of the circuit $C$ are constructed so that they are only satisfied by valid tuples $(x, y)$, i.e., by tuples where $y = f(x)$ actually holds.
This allows the verifier to accept the proof. The witness $W$ can be considered a trace of the computation under consideration, and constraints, i.e., the relation $R$, are assertions that enforce the correctness of the computation.
In the provided example, the witness will contain both $x$, $y$.

Figure~\ref{fig:montgomery-circuit} contains the program for Montgomery curve addition in Circom (one of the most popular DSL for writing ZKP circuits).
Executing its assignment instructions (i.e., computation obtained by \texttt{<-{}-} or \texttt{<==}) on two example curve points yields the field elements shown in Figure~\ref{fig:montgomery-witness}, which compose \emph{witness}.
Each constraint in Figure~\ref{fig:montgomery-constraints} is obtained from an \texttt{===} or \texttt{<==} in Figure~\ref{fig:montgomery-circuit}. Substituting the numbers from Figure~\ref{fig:montgomery-witness} into those three equations makes every left-hand side equal 0, hence the tuple
\(\bigl(\texttt{in1},\texttt{in2},\texttt{out}\bigr)\) satisfies the circuit and can be turned into a valid proof.

In practice, developers use a DSL to write high-level code that describes which parts of the computation must be constrained and how.
A compiler for the DSL then produces polynomial equations describing the constraints. Certain computations can be kept \emph{out of the circuit}, with constraints merely validating that the external steps have been performed correctly. This approach can substantially reduce proof-generation costs. Ultimately, the circuit ensures that the verifier can check, with minimal overhead, that the prover indeed holds a valid witness for the asserted statement, all while preserving the zero-knowledge property.

\begin{figure}[t]
  \begin{center}
    \includegraphics[width=0.42\textwidth]{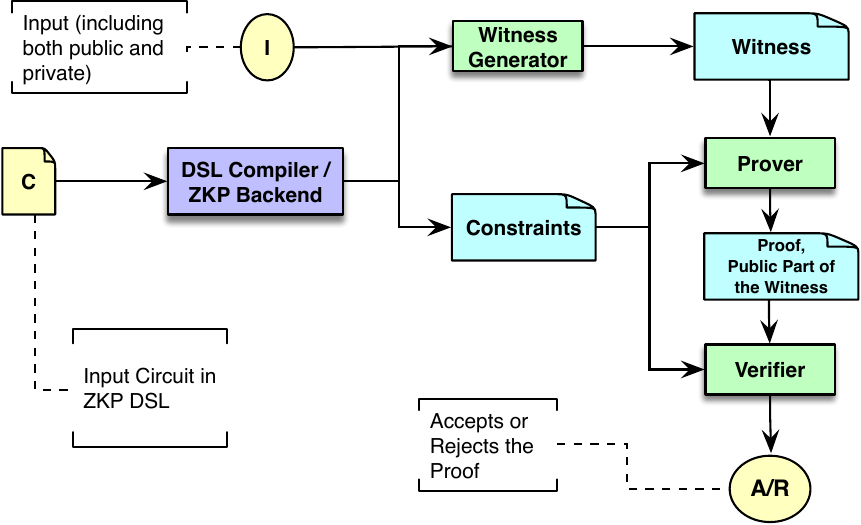}
  \end{center}
  \caption{Simplified compilation and proof generation pipeline of ZKP circuits}
    \label{fig:overview}
\end{figure}

Figure~\ref{fig:multiplierSafe} shows a small circuit expressed in \emph{Circom}~\cite{belles2022circom}, a popular DSL for expressing ZK circuits (such languages are known as zkDSLs). Given this circuit, the Circom compiler produces both (i) a \emph{witness generator}, which calculates \lstinline{c} from \lstinline{a}, \lstinline{b}, i.e., produces the witness to be proved; and (ii) a \emph{constraint system} ensuring any valid proof satisfies \(\texttt{c} = \texttt{a}\times\texttt{b}\). Then, a SNARK backend provides a \emph{prover} (which uses the witness to produce a proof) and a \emph{verifier} (which effectively checks a produced proof against the associated constraints and public inputs).
Figure~\ref{fig:overview} depicts the pipeline for compiling a ZKP circuit and using it to generate and verify a proof.

\textbf{Different Approaches to ZKP Development.}
There are three different ways that developers can use ZKPs.
\emph{(i) zkDSLs} such as Circom allow developers to write the circuit that implements their specifications directly. This offers tight control over efficiency but demands significant low-level expertise. \emph{(ii) Domain-specific transpilers} allow developers to write higher-level statements (e.g., a regex in \texttt{zk-regex}~\cite{zk-regex}) which are then transpiled into circuits (typically expressed in a zkDSL). While a good fit for specific tasks, this approach is not general-purpose and can introduce hidden inefficiencies and bugs if the transpiler is not correct.
\emph{(iii) ZK Virtual Machines (zkVMs)} like Cairo~\cite{goldberg2021cairo} or RISC0~\cite{bruestle2023riscZeroZkVM} encode an entire CPU-like execution loop into some large circuits, allowing the execution of regular programs (e.g., in Rust) to be proved correct. This generality comes at the cost of performance. Notably, in all of these approaches, circuits remain a core component; what varies is whether they are written directly by a developer, or created by upstream tooling.

\begin{table*}[t]
    \captionof{table}{Overview of ZKP Bug Categories}
    \label{tab:zkp_bug_categories}
    \begin{footnotesize}
    \begin{tabular}{lp{4.5cm}p{6.2cm}}
        \toprule
        \textbf{Category} & \textbf{Manifestation} & \textbf{Detection Approach} \\
        \midrule
        \textbf{Completeness} & Valid inputs fail to produce valid proofs & Use known valid inputs from a spec or reference; if a proof is not generated, there is a bug\\ 
        \hdashline
        \textbf{Correctness}  & Valid or invalid proof that contradicts intended behavior & Compare circuit outputs to expected outputs (via reference impl./formal spec) \\ 
        \hdashline
        \textbf{Soundness}    & Dishonest prover can produce invalid proofs that the verifier accepts & Requires generating invalid witnesses that “should fail,” but does not fail \\
        \bottomrule
    \end{tabular}
    \end{footnotesize}
\end{table*}

\textbf{Common Vulnerabilities.}
Implementing ZKP circuits is error-prone, and various kinds of bugs can undermine the key ZKP properties of completeness, soundness and zero-knowledge.

A critical class of bugs are \textbf{soundness} issues that make a circuit \textit{under-constrained}, such that witnesses are not fully constrained. This situation typically violates intended functionality, and allows a malicious prover to choose a bogus witness that satisfies all the given constraints but does not correspond to a trace that the circuit can actually produce. 
A simple example is shown in Figure~\ref{fig:multiplierSoundness}, a variant of Figure~\ref{fig:multiplierSafe} where the result of the multiplication is unconstrained, so that the prover can convince the verifier that the circuit was used to determine bogus multiplicative relationships between numbers (e.g., that \texttt{2*5=100}).
Figure~\ref{fig:montgomeryoverall} shows a more realistic example.\footnote{This example was extracted from the following report: \url{https://veridise.com/wp-content/uploads/2023/02/VAR-circom-bigint.pdf}.} \texttt{MontgomeryAdd}~\cite{montgomeryadd} is used when adding two points on a Montgomery elliptic curve. Given two points \texttt{(in1[0], in1[1])} and \texttt{(in2[0], in2[1])} on the curve, a new point \texttt{(out[0], out[1])} is calculated that effectively represents their sum under elliptic-curve arithmetic.
Although the computation of the circuit of Figure~\ref{fig:montgomery-circuit} is correct, this circuit is under-constrained because an implicit assumption that $in2[0] - in1[0] \neq 0$.
Thus, a malicious prover can convince a verifier that the addition of two points can be another value than the expected one because of the freedom to control \texttt{lambda}. For example, consider the scenario where all inputs are $0$, then the prover can pick any value for \texttt{lambda} and compute the output accordingly, regardless of whether the value constitutes a correct result from the circuit. Note that the verifier will accept the proof produced because the constraints will be satisfied.

In contrast, an \textit{over-constrained} circuit contains extraneous constraints that make the constraint system too restrictive for some legitimate inputs.
Such bugs lead to \textbf{completeness} issues.
Figure~\ref{fig:multiplierCompleteness} presents a variant of the multiplier circuit that suffers from a completeness bug. 
While the circuit works correctly when both $a$ and $b$ are $0$ or $2$, for other values, the constraints prevent the prover from producing a valid witness, resulting in failure to generate a proof.

Finally, \textbf{correctness} errors occur when the circuit fails to implement the intended behavior. For instance, as shown in Figure~\ref{fig:multiplierCorrectness}, the circuit uses addition instead of multiplication to compute the product. Even if the witness satisfies the circuit's constraints, the output does not reflect the correct computation, causing either valid proofs to be rejected or invalid proofs to be accepted.

Table~\ref{tab:zkp_bug_categories} outlines the different bug categories. Note that in contrast to prior work~\cite{chaliasos2024sok}, we separate correctness and soundness issues, as they manifest differently.

\section{Challenges, Solutions and Open Problems of Fuzzing ZKP Programs}
\label{sec:fuzzing}

Applying fuzzing techniques to ZKP circuits presents unique challenges arising from their specialized nature, including uncommon bug manifestations and the separation between constraints and computation. We now discuss the main challenges in fuzzing ZKP circuits and propose potential solutions. 

\subsection{Key Challenges}

\textbf{Test Oracle Problem.}
A critical challenge in fuzzing is the \emph{test oracle problem}. In conventional software testing, a well-defined oracle indicates whether an output is correct for a given input~\cite{barr2014oracle}. For ZKPs, the objective is to define oracles that can detect completeness, correctness, and soundness bugs.

\emph{Completeness} bugs are relatively straightforward to detect by verifying that valid inputs produce valid proofs, i.e., that unexpected failures are not observed.
The completeness bug of Figure~\ref{fig:multiplierCompleteness} could be detected by trying any inputs beyond $0$ and $2$.
Identifying \emph{correctness} bugs requires knowledge of the expected behavior, as they manifest when an honest prover outputs a proof that, while valid in the context of the circuit, conflicts with the expected behavior.
Detecting the correctness bug of Figure~\ref{fig:multiplierCorrectness} would require knowing that the circuit under test needs to perform multiplication instead of addition.
\emph{Soundness} bugs are harder to detect because they involve \textit{invalid} proofs that the verifier erroneously accepts. Missing or insufficient constraints commonly cause such errors, as developers are working with low-level DSLs, a different programming model, and the need for optimizations.
To detect the soundness bug of Figure~\ref{fig:multiplierSoundness}, one would need to generate a witness that satisfies the constraints but does not match the logic of the computation.

\textbf{Input/Witness Generation.}
Input (or \emph{witness}) generation aligns closely with oracle design. For completeness and correctness bugs, producing valid inputs is relatively simple if a reference or specification exists. Generating invalid-yet-accepted witnesses (for soundness tests), however, is significantly harder: 
\emph{(i)} randomly mutating a valid input often yields a proof rejected outright by the verifier; 
\emph{(ii)} determining whether a mutated witness is 
\emph{incorrect} yet still accepted requires advanced oracles, reference checks, or carefully designed invariants; and 
\emph{(iii)} a small local mutation can invalidate other parts of the circuit unless it is carefully guided.

\textbf{Efficiency and Performance Overheads.}
ZKP proof generation and verification entail computationally expensive operations. For example, proving could take up to hours for large real-world circuits. Fuzzing usually involves running large quantities of tests, but each proof can take significant time. This overhead renders standard fuzzing techniques infeasible in some cases.

\textbf{Support for Multiple DSLs and Frameworks.}
ZKP circuits can be written in various DSLs (Circom, Noir, ZoKrates, Halo2, etc.), each mapping to potentially different proof systems. A fuzzing tool for one DSL might not port cleanly to another, leading to fragmented solutions and additional engineering challenges. Working on the arithmetization level (e.g., R1CS) might be possible, but critical information such as circuit separation is not preserved at this level. While not ZK-specific, this problem is magnified in the ZK space due to the numerous similar DSLs available to developers.

\begin{figure}[t]
    \centering
    \includegraphics[scale=0.42]{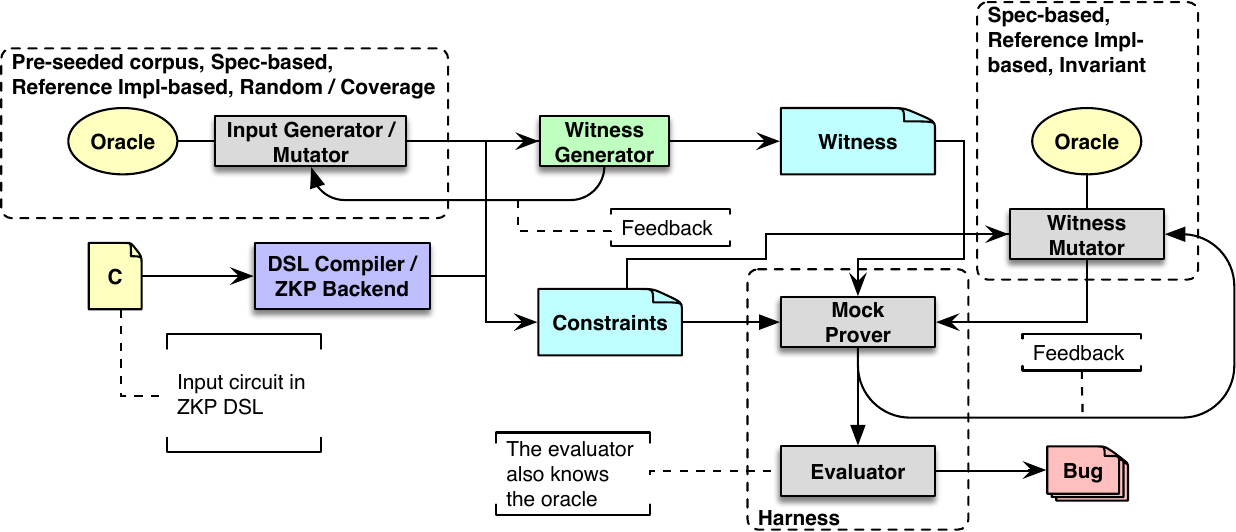}
    \caption{Fuzzing ZKP circuits}
    \label{fig:fuzzing}
\end{figure}

\textbf{Working with Cryptographic Primitives.}
Many ZKP circuits integrate cryptographic operations such as hashes or signatures. Random mutations rarely reveal subtle flaws (e.g., partial collisions or bit misalignment). Moreover, cryptographic primitives often impose strict input conditions, such as specific preimages that should result to a hash after some operations, that are extremely unlikely to be met by randomly mutated inputs. This challenge is similar to issues encountered in fuzzing checksum-based systems~\cite{wang2010taintscope}, where valid inputs that satisfy the checksum constraints are rare, thereby hindering the exploration of critical circuit paths.

\subsection{Proposed Solutions}
\label{sec:solutions}
We discuss potential solutions to the challenges outlined in the previous section. 
Figure~\ref{fig:fuzzing} outlines the primary components required to fuzz test ZKP circuits.
The first step is to compile the circuit (i.e., the system under test) into a constraint system along with a witness generator. Next, an input generator produces candidate inputs based on an oracle (such as a specification), and these inputs are used to generate the corresponding witnesses. The witnesses are then processed by a mock prover, and an evaluator checks whether the results meet the oracle’s expectations, thereby reporting any bugs. Furthermore, a witness mutator can further mutate these witnesses to produce variants that may reveal soundness issues. Finally, feedback from both the input generation/mutation and witness mutation stages is used to guide subsequent mutations.

\begin{table}[t]
\caption{Comparison of oracle strategies; $^*$ indicates that novel invariants can handle more bug types}
\label{tab:oracle_strategy_comparison}
\resizebox{\linewidth}{!}{
\begin{tabular}{lllll}
\toprule
\textbf{Oracle Strategy} & \textbf{Prerequisites} & \textbf{Bug Types Detected} \\
\midrule
\textbf{Spec-Based} 
  & Specification 
  & Completeness, Correctness, Soundness \\ 

\textbf{Differential} 
  & Reference Implementation 
  & Completeness, Correctness, Soundness \\

\textbf{Invariants} 
  & None
  & Soundness$^*$ \\
\bottomrule
\end{tabular}
}
\end{table}

\textbf{Oracle Design Strategies.}
When a (machine-readable) specification or reference implementation is available, (in)correct-by-construction input generation or differential testing~\cite{mckeeman1998differential} can resolve the oracle problem. If neither exists, one might leverage generic invariants (for instance, \textit{deterministic outputs for the same inputs}~\cite{pailoor2023automated}) to detect specific bug categories. Table~\ref{tab:oracle_strategy_comparison} summarizes the bug types that each oracle can detect. Although common oracles such as crashes can be used, those would not be able to detect the majority of ZKP-specific bugs.

\begin{figure*}[t]
     \centering
     \begin{subfigure}{0.43\textwidth}
         \centering
         \includegraphics[scale=0.45]{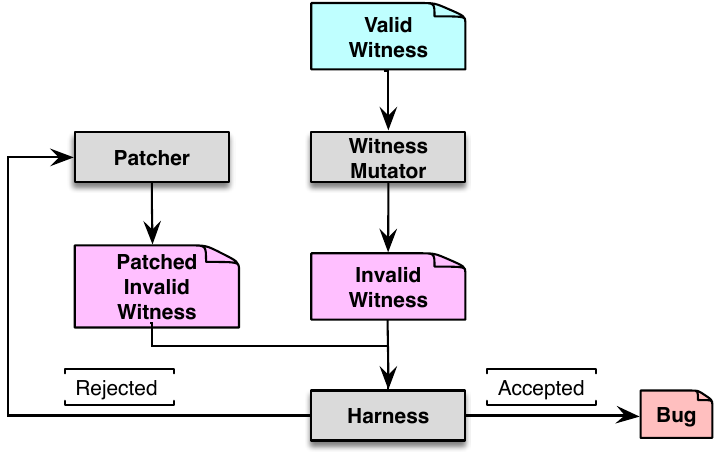}
         \caption{Mutation-based approach}
         \label{fig:mutation}
     \end{subfigure}
     \begin{subfigure}{0.37\textwidth}
         \centering
         \includegraphics[scale=0.45]{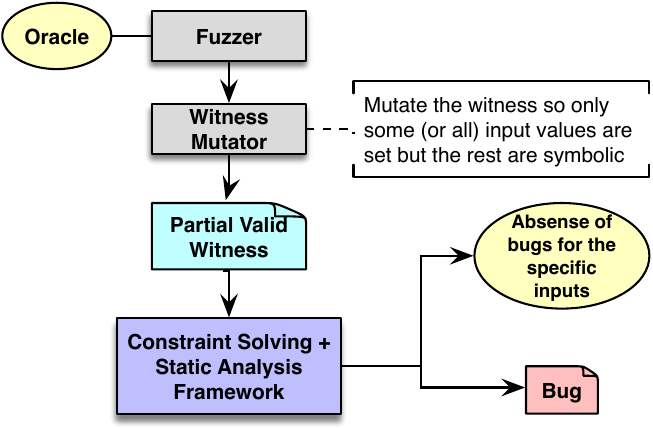}
         \caption{Hybrid approach}
         \label{fig:hybrid}
     \end{subfigure}
     \caption{
     Workflow of a mutation-based approach for finding soundness bugs, and a hybrid approach that can be used to improve symbolic and static analyses to detect soundness bugs
     }
     \label{fig:mutation_hybrid}
\end{figure*}

\textbf{Input Generation for Completeness/Correctness vs. Soundness.}
Unlike completeness and correctness bugs,
which can be caught by feeding valid or invalid inputs into a circuit and checking whether proofs and results match expectations, soundness bugs require crafting \textit{witnesses} that the circuit ought to reject but instead (incorrectly) accepts. Randomized invalid witness generation is unlikely to be effective for non-trivial circuits because bugs typically occur in corner cases, e.g., in Figure~\ref{fig:montgomery-circuit}, the bug only manifests when \texttt{in2[0] = in1[0]}. That leaves the following options:

\emph{(i) Mutation From Valid Witnesses}: A promising approach is to begin with a valid witness and then introduce slight modifications that make it invalid according to the oracle, thus exposing under-constrained soundness issues (Figure~\ref{fig:mutation}). An additional improvement is to keep mutations local: if an error is introduced, adjust
(i.e., patch) the remaining parts of the witness to ensure only some constraints are targeted. The intuition is that some constraints might be missing or incorrect, while others may still be valid; if the mutated value conflicts with subsequent constraints and the witness is not patched, the mismatch will cause a failure there, obscuring the bug.

\emph{(ii) Hybrid Methods}: Symbolic and static analysis has been used to detect soundness bugs in ZKP circuits~\cite{pailoor2023automated,wen2023practical}, but suffers from scalability or precision issues.
We envisage a hybrid approach where fuzzing is used to find \textit{good inputs},
after which the other techniques to reason about the properties of a circuit based on these inputs; see Figure~\ref{fig:hybrid}.
This will not provide any guarantees, but it could be a pragmatic approach to detecting bugs. The main intuition for this approach is that it will drastically reduce search space.

\textbf{Optimizing Performance for ZKP Fuzzing.}
Because proof generation is expensive, avoiding cryptographic steps on every iteration is crucial. Pre-checking constraints in a `mock' mode can save time if the DSL supports it (e.g., MockProver in Halo2). It is critical for DSLs to add support for mock proving to make fuzzing circuits possible. Additionally, sub-circuit fuzzing might be beneficial when having to fuzz test huge circuits. Yet, the oracle should be adapted because there might be a `bug' (e.g., a missing constraint) in the sub-circuit, but other constraints in other circuits or the main circuit could include the missing check. Hence, when sub-circuits are targeted, the findings must be validated against the entire circuit to prevent false positives.

\textbf{Cross-DSL Compatibility.}
A generic approach might unify circuits in a common intermediate representation (IR). However, any discrepancies between the DSL’s compiler and the IR layer can introduce false positives or negatives (e.g., due to compiler bugs). Still, an IR-based framework potentially allows a single fuzzer to test multiple DSLs. A framework such as CirC~\cite{ozdemir2022circ} could be helpful for this use case. Although this is not a ZK-specific issue, it is even more critical in the ZK space due to the fragmentation because of the different DSLs, which more or less allow the representation of similar programs and primitives.

\textbf{Handling Cryptographic Primitives.}
A specialized oracle or white-box hooking of cryptographic components can help validate these sub-circuits. If the fuzzing target is a cryptographic function, domain-specific checks or differential testing against a reference implementation may suffice. The main difficulty is ensuring the fuzzer explores cryptographic logic deeply. For example, if a circuit takes as inputs a preimage of a hash and the hash value, we need to guide the fuzzer on how to compute the hash for a random preimage to explore the whole circuit (given that it has more functionality). 

\subsection{Open Problems and Future Directions}

\textbf{Novel Invariants as Oracles.} In Section~\ref{sec:solutions} we discussed how an existing invariant~\cite{pailoor2023automated} can be employed to detect soundness bugs with fuzzing. More complex invariants capturing deeper properties of ZKP circuits could enable the detection of additional classes of bugs in the absence of a full specification or reference implementation.

\textbf{Practical Patching Mechanisms for Soundness Bugs.} Although mutation-based approaches can expose under-constrained circuits, efficiently ``patching'' the rest of the witness to isolate targeted constraints remains complex. Implementations for dynamic witness adaptation could significantly improve the discovery of subtle soundness bugs.


\textbf{Improved Coverage Metrics for ZKP Circuits.} Traditional code-based coverage metrics do not directly apply to constraint-based logic. Developing customized coverage strategies (e.g., tracking how many constraints are actively `exercised' in each test) may guide fuzzers to explore corner cases more systematically.

\textbf{Combining Fuzzing with Static and Symbolic Analyses.} Hybrid approaches that couple lightweight static checks or partial symbolic exploration with mutation-based fuzzing could amplify bug-finding capabilities. Integrating these methods seamlessly in the context of ZKP circuits remains an open challenge but offers high potential to bolster both coverage and depth of analysis.

\medskip
In summary, fuzzing ZKP circuits is still in its infancy, leaving significant avenues for future research. Advancing this field is crucial to ensure that ZKPs remain secure and reliable as they continue to underpin an expanding range of modern applications.

\section{Case Study: Fuzzing ZK-Email}\label{sec:casestudy}

\begin{figure*}[t]
    \centering
    \includegraphics[scale=0.23]{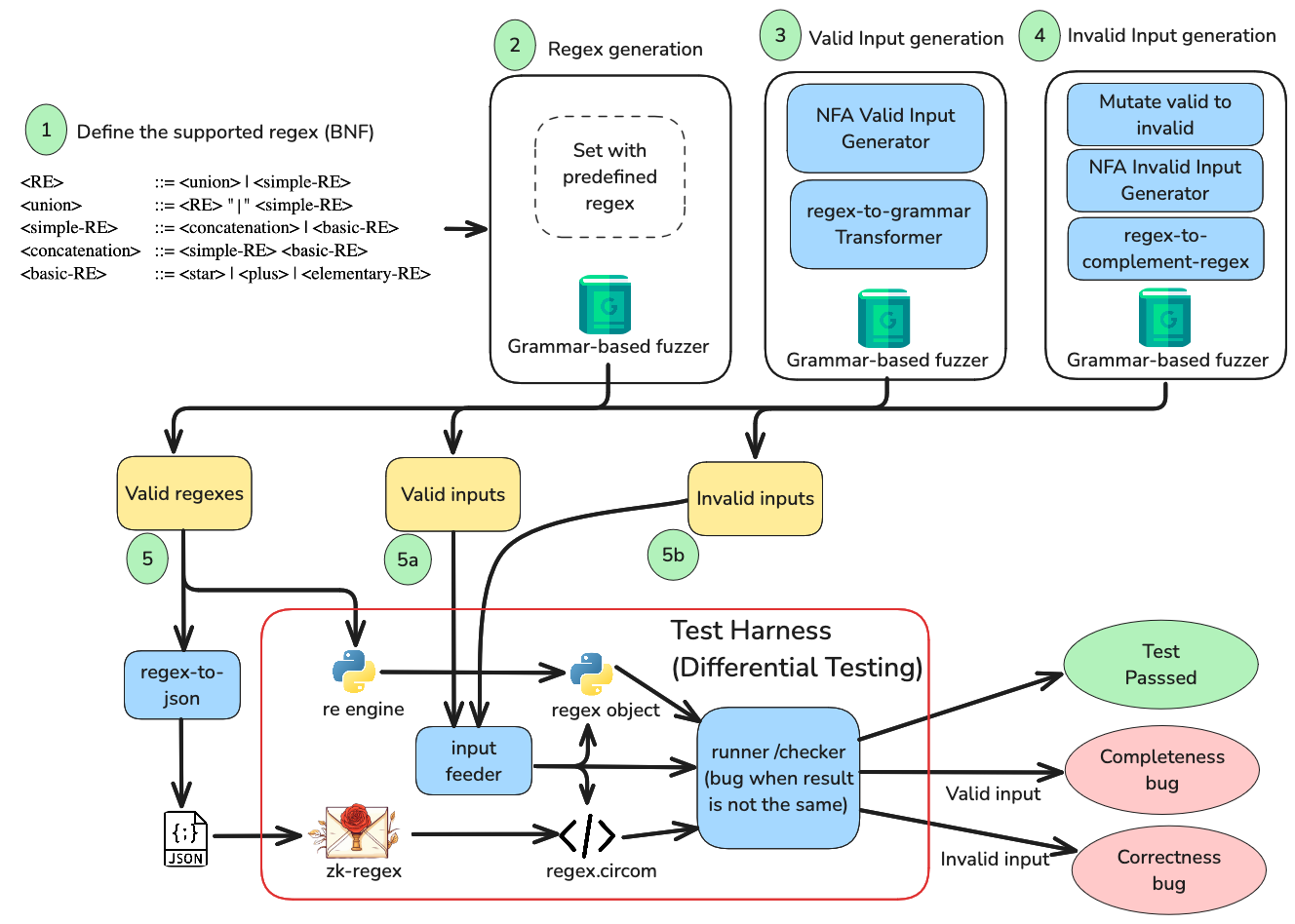}
    \caption{Fuzzing zk-regex}
    \label{fig:fuzzing-zkregex}
\end{figure*}

Zk-email is a core library that provides an easy way to prove properties about emails within a circuit~\cite{zk-regex}. One common use case is to prove ownership of an account (e.g., Twitter account) by demonstrating that a user received an email from Twitter containing their username. This is achieved by verifying the DKIM signatures and proving the inclusion of the username in the email body within a circuit. Zk-email relies on regular expressions for substring matching, and to facilitate this, it provides a transpiler that converts a regex into a Circom or Noir circuit. As a core component used by many production-level ZK applications, the security of zk-regex is critical.

To fuzz-test the circuits produced by \texttt{zk-regex}, we address the test oracle problem by using a reference implementation of Python's regex module combined with differential testing. For input generation, we employ correct-by-construction and incorrect-by-construction inputs based on a specification of what constitutes a valid regex and the expected behavior of a given regex. 
Figure~\ref{fig:fuzzing-zkregex} depicts the end-to-end workflow of our fuzzer. We begin by writing a BNF specification of the fragment of regular-expression syntax supported by \texttt{zk-regex}. Then we employ a grammar-based fuzzer along with a pre-seeded corpus of important regexes, to get valid regexes. For each regex we then construct two types of test strings: \emph{valid} inputs, which should be accepted, are obtained both by grammar expansion of the regex by employing a grammar-based fuzzer and by enumerating accepting paths in the corresponding NFA, whereas \emph{invalid} inputs, which should be rejected, are generated through random mutation of valid inputs, exploration of non-accepting NFA branches, and generation from the grammar of the regex’s complement. Next, each regex is wrapped in a JSON expected by \texttt{zk-regex}, compiled to a Circom circuit, and executed by using valid and invalid inputs. The harness performs differential testing between Python’s \texttt{re} module and that of the generated circuit, and any disagreement is reported as a bug based on the used oracle that depends on if the input is valid or invalid.

Our campaign was conducted over one week, during which we fuzzed the library for one hour per day on a 16-core machine, continuously enhancing our fuzzer and incorporating fixes for some of the reported bugs in \texttt{zk-regex}. Our approach detected completeness and correctness bugs, although it did not uncover soundness bugs since we did not fuzz the witness. Despite the simplicity of our method, we revealed a total of 13 new bugs (5 correctness and 8 completeness), underscoring that even basic fuzzing campaigns can uncover significant issues in ZKP circuits. Notably, all $13$ bugs were confirmed by the developers of \texttt{zk-regex}. We leave as future work the implementation and evaluation of the additional techniques described in Section~\ref{sec:fuzzing}.

\section{Related Work}
\point{ZKP Security and Automated Analysis}
Recent studies have categorized ZKP bugs and threats to different layers, highlighting a comprehensive taxonomy of vulnerabilities~\cite{chaliasos2024sok}. Several works have also focused on attacking ZKP constructions, demonstrating how oversights can lead to severe security vulnerabilities in all ZKP implementations that use the specific constructions~\cite{dao2023weak,khovratovich2025prove,ciobotaru2024last}. Another line of work has focused on rigorous verification of ZKP components, for instance, verifying the verifier itself to ensure correctness~\cite{firsov2024ouroboros}. While formal verification has been used to prove the absence of bugs in ZKP circuits~\cite{pailoor2023automated,coglio2023compositional,ozdemir2023bounded,miguel2024scalable}, it often faces scalability challenges. Similarly, static analysis approaches~\cite{wen2023practical,soureshjani2023automated} can detect certain classes of issues but may suffer from precision limitations and heuristics that miss broader categories of bugs. These hurdles underscore the need for complementary, scalable, and precise techniques, such as fuzzing, that could reveal critical bugs.
\point{Fuzzing ZKP Implementations}
Fuzzing ZKP implementations has gained traction in a few domain-specific contexts. For instance, recent work identified completeness bugs in a specific zkEVM implementation by systematically mutating and monitoring execution paths~\cite{li2024famulet}. Further, metamorphic testing has been employed to uncover errors in ZKP compilers~\cite{hochrainer2024fuzzing,xiao2025mtzk}. Finally, SnarkProbe~\cite{fan2024snarkprobe} focuses on revealing bugs in ZKP DSL backends. These papers either focus on non-circuit bugs or struggle with the challenges we outlined in Section~\ref{sec:fuzzing}, such as solving the oracle problem for soundness bugs.

\section{Conclusion}
We have discussed the unique challenges of fuzzing ZKP circuits, including issues with test oracles, input/witness generation, and performance overheads,
and sketched practical solutions such as differential testing, invariant-based oracles, and hybrid methods to enhance bug detection. Our zk-regex case study demonstrates that even a simple fuzzing campaign can reveal significant bugs, highlighting the potential of fuzzing to improve ZKP security.
The remaining open problems we have identified provide fruitful ideas for future fuzzing research in this domain.
%

\begin{acks}
We thank the anonymous reviewers for their constructive feedback. Further, we would like to thank Shreyas Londhe and Aayush Gupta from zkEmail for firmly confirming our bugs and brainstorming potential extensions of our fuzzing framework for zk-regex. We also want to thank Kostas Ferles, Burak Kadron, and Ben Mariano from Veridise Inc., for fruitful discussions on hybrid fuzzing techniques for ZK circuits. This work has been partially funded by Aztec Labs.
\end{acks}

\bibliographystyle{ACM-Reference-Format}
\bibliography{main}

\end{document}